\documentstyle[11pt,axodraw]{article}
\begin{document}
\setcounter{footnote}{0}
\renewcommand{\thefootnote}{\alph{footnote}}
\renewcommand{\theequation}{\thesection.\arabic{equation}}
\newcounter{saveeqn}
\newcommand{\add}{\addtocounter{equation}{1}}
\newcommand{\alpheqn}{\setcounter{saveeqn}{\value{equation}}%
\setcounter{equation}{0}%
\renewcommand{\theequation}{\mbox{\thesection.\arabic{saveeqn}{\alph{equation}}}}}
\newcommand{\reseteqn}{\setcounter{equation}{\value{saveeqn}}%
\renewcommand{\theequation}{\thesection.\arabic{equation}}}
\newenvironment{nedalph}{\add\alpheqn\begin{eqnarray}}{\end{eqnarray}\reseteqn}
\newsavebox{\PSLASH}
\sbox{\PSLASH}{$p$\hspace{-1.8mm}/}
\newcommand{\PS}{\usebox{\PSLASH}}
\newsavebox{\PARTIALSLASH}
\sbox{\PARTIALSLASH}{$\partial$\hspace{-2.3mm}/}
\newcommand{\PARTIALS}{\usebox{\PARTIALSLASH}}
\newsavebox{\ASLASH}
\sbox{\ASLASH}{$A$\hspace{-2.1mm}/}
\newcommand{\AS}{\usebox{\ASLASH}}
\newsavebox{\KSLASH}
\sbox{\KSLASH}{$k$\hspace{-1.8mm}/}
\newcommand{\KS}{\usebox{\KSLASH}}
\newsavebox{\LSLASH}
\sbox{\LSLASH}{$\ell$\hspace{-1.8mm}/}
\newcommand{\LS}{\usebox{\LSLASH}}
\newsavebox{\QSLASH}
\sbox{\QSLASH}{$q$\hspace{-1.8mm}/}
\newcommand{\QS}{\usebox{\QSLASH}}
\newsavebox{\DSLASH}
\sbox{\DSLASH}{$D$\hspace{-2.8mm}/}
\newcommand{\DS}{\usebox{\DSLASH}}
\newsavebox{\DbfSLASH}
\sbox{\DbfSLASH}{${\mathbf D}$\hspace{-2.8mm}/}
\newcommand{\DBFS}{\usebox{\DbfSLASH}}
\newsavebox{\DELVECRIGHT}
\sbox{\DELVECRIGHT}{$\stackrel{\rightarrow}{\partial}$}
\newcommand{\PARVECR}{\usebox{\DELVECRIGHT}}
\thispagestyle{empty}
\begin{flushright}
IPM/P-2002/020
\par
hep-th/0206137
\end{flushright}
\vspace{0.5cm}
\begin{center}
{\Large\bf{On the $\beta$-Function and Conformal Anomaly
 of Noncommutative QED with Adjoint Matter Fields}}\\
\vspace{1cm} {\bf N\'eda Sadooghi$\
^{\dagger,}$}\footnote{\normalsize{Electronic address:
sadooghi@theory.ipm.ac.ir}} \hspace{0.2cm} and \hspace{0.2cm}{\bf
Mojtaba Mohammadi$\ ^{\ddagger,}$}\footnote{\normalsize{Electronic
address:
m.mohammadi@mehr.sharif.edu}}  \\
\vspace{0.5cm}
{\sl ${\ ^{\dagger, \ddagger}}$ Department of Physics, Sharif University of Technology}\\
{\sl P.O. Box 11365-9161, Tehran-Iran}\\
and\\
{\sl ${\ ^{\dagger}}$ Institute for Studies in Theoretical Physics and Mathematics (IPM)}\\
{\sl{School of Physics, P.O. Box 19395-5531, Tehran-Iran}}\\
\end{center}
\vspace{0cm}
\begin{center}
{\bf {Abstract}}
\end{center}
\begin{quote}
In the first part of this work,
a perturbative analysis up to one-loop order is carried out to
determine the one-loop $\beta$-function of noncommutative $U(1)$ gauge
theory with matter fields in the adjoint representation. In the second part, the
conformal anomaly of the same theory is calculated using the
Fujikawa's path integral method. The value of the one-loop $\beta$-function
calculated in both methods coincides. As it turns out, noncommutative QED with matter 
fields in the adjoint representation is asymptotically free for the number of flavor degrees of freedom $N_{f}<3$.
\end{quote}
\hspace{0.8cm}
\newpage
\setcounter{page}{1}
\section{Introduction}
\setcounter{section}{1} In the past few years, the noncommutative
gauge theories have attracted much attention and are studied
seriously by many authors \cite{review}. They are interesting due
to their realization in the String Theory. In the decoupling
limit, noncommutative gauge theories can occur in the world volume
of D-branes in the presence of the constant background
$B_{\mu\nu}$-field \cite{string}. But, noncommutative gauge
theories can also be regarded as nontrivial deformation of the
ordinary commutative gauge theories. This deformation is simply
realized by replacing the ordinary commutative product of
functions by the noncommutative $\star$-product:
\begin{eqnarray*}
f\left(x\right)\star g\left(x\right)\equiv
e^{\frac{i\theta_{\mu\nu}}{2}\ \frac{\partial}{\partial\xi_{\mu}}\
\frac{\partial}{\partial\zeta_{\nu}}
}f\left(x+\xi\right)g\left(x+\zeta\right)\bigg|_{\xi=\zeta=0},
\end{eqnarray*}
where $\theta_{\mu\nu}$ is a real, and antisymmetric tensor,
reflecting the noncommutativity of the space-time coordinates.
\par
The most natural candidate for such noncommutative gauge theories
is the noncommutative QED,
where the coupling of the matter fields to the gauge fields can be
given in the fundamental, antifundamental or adjoint
representations.
\par\noindent
Different aspects of noncommutative $U(1)$ gauge theory with {\it
{fundamental}} matters are studied intensively in the literature, and
various results are achieved
[for a review, see Ref. \cite{review} and the references therein].
 A perturbative analysis of the one-loop
$\beta$-function of this theory shows, that for the
number of flavor degrees of freedom $N_{f}<6$, the NC-$U(1)$ gauge
theory is asymptotically free \cite{hayakawa}. Asymptotically free
theories appear also in the ordinary commutative non-Abelian gauge
theories. Commutative $SU(3)$, {\it e.g.}, exhibits asymptotic
freedom for $N_{f}\leq 16$.
\par
Recently, the same one-loop $\beta$-function of noncommutative QED
with {\it {fundamental}} matter fields is also calculated by
studying the relation between the conformal anomaly of
noncommutative QED and the $\beta$-function of the theory [see
Ref. \cite{nakajima}]. As is known, noncommutative gauge theories
include the noncommutativity parameter $\theta$ of dimension
[length]$^2$. It is therefore expected, that the Weyl symmetry of
noncommutative gauge theories is broken at classical level, even in
the massless limit. But, as in the ordinary commutative case, the
Weyl symmetry of noncommutative $U(1)$ is also broken by quantum
corrections. This is defined as the conformal anomaly of
noncommutative QED. These quantum corrections can be calculated
using, {\it e.g.}, the Fujikawa's path integral method
\cite{fujikawa}, which is also used to calculate the global and
gauge anomalies of noncommutaive $U(1)$ and $U(N)$ gauge theories
\cite{axial}. As it is shown in Ref. \cite{nakajima}, the value of 
the one-loop
$\beta$-function of noncommutative QED with fundamental matters
 calculated from the conformal anomaly coincides with the
result of perturbative analysis from Ref. \cite{hayakawa}.
\par
What concerns the noncommutative QED with {\it adjoint}
matter field coupling, there are good reasons to study these
theories too. As it is argued by Seiberg in Ref. \cite{seiberg},
noncommutative Yang-Mills theory appears as a manifestation of a
matrix model and the underlying model forces the coupling of the
matter to the gauge fields to be in the adjoint representation of
noncommutative algebra. Noncommutative QED with matter fields in
the adjoint representation is first studied in Ref.
\cite{susskind}, where the one-loop vacuum polarization tensor of
the theory was calculated in the framework of perturbation theory.
It is shown, that in noncommutative gauge theories with matter
fields in the adjoint representations, nonplanar Feynman diagrams
appear which exhibit the important UV/IR mixing phenomena of
noncommutative Field Theory \cite{minwalla}. The study of
noncommutative gauge theories with adjoint matter fields is also
interesting from the point of view of noncommutative standard
model and a possible $\theta$-dependent correction to the neutrino
mass \cite{sheikh, wess}.
\par
In this letter, the one-loop $\beta$-function of NC-$U(1)$ gauge
theory with adjoint matter fields will be calculated by making use
of two different methods. We will first introduce the
noncommutative QED with adjoint matter fields and its
corresponding Feynman rules in Sect. 2. Then, in Sect. 3, a
perturbative analysis up to one-loop order will be performed to
calculate the one-loop $\beta$-function of the theory. In Sect. 4,
the conformal anomaly of noncommutative QED with adjoint matter
fields will be calculated using the Fujikawa's path integral
method \cite{fujikawa}. As in the fundamental case, the value of
the one-loop $\beta$-function coincides in both methods. We will
show that, noncommutative QED with adjoint matter fields is
asymptotically free for $N_{f}<3$. Sect. 5, is devoted to
discussions.
\section{Noncommutative QED with Adjoint Matter Fields}
The action of noncommutative QED with matter fields in the adjoint
representation is given by:
\begin{eqnarray}\label{M1}
S[\psi,\overline{\psi};A_{\mu}]=S_{\small{gauge}}[A_{\mu}]+
S_{{\small{matter}}}[\psi,\overline{\psi}]+S_{{\small{ghost}}}[c,\overline{c}],
\end{eqnarray}
with the gauge part
\begin{eqnarray}\label{M2}
S_{{\small{gauge}}}[A_{\mu}]=-\frac{1}{4}\int d^{D}x\
F_{\mu\nu}(x)\star F^{\mu\nu}(x).
\end{eqnarray}
The field strength tensor is defined by:
\begin{eqnarray}\label{M6}
F_{\mu\nu}(x)=\partial_{\mu}A_{\nu}(x)-\partial_{\nu}A_{\mu}(x)-ig[A_{\mu}(x),A_{\nu}(x)]_{\star}.
\end{eqnarray}
The matter field part of the action is given by
\begin{eqnarray}\label{M3}
S_{{\small{matter}}}[\psi,\overline{\psi}]=\int
d^{D}x\bigg[i\overline{\psi}(x)\gamma^{\mu}\star
D_{\mu}\psi(x)-m\overline{\psi}(x)\star\psi(x)\bigg],
\end{eqnarray}
with the covariant derivative
\begin{eqnarray}\label{M4}
D_{\mu}\psi(x)=\partial_{\mu}\psi(x)-ig\big[A_{\mu}(x),\psi(x)\big]_{\star}.
\end{eqnarray}
Finally the ghost part of the action reads
\begin{eqnarray}\label{M5}
S_{\small{ghost}}= \int d^{D}x\
\bar{c}(x)\star(-\partial_{\mu}D^{\mu})c(x),
\end{eqnarray}
where $c(x)$ are the ghost fields in the adjoint representation.
The covariant derivative in Eq. (\ref{M5}) is defined by the
derivative from  Eq. (\ref{M4}). The Feynman rules of
noncommutative QED with adjoint matter fields can be calculated
from the above actions and read [see also Ref. \cite{susskind}]:
\vskip0.3cm\noindent {\it i) Fermion Propagator:}
\begin{eqnarray}\label{FX1a}
\SetScale{1}
    \begin{picture}(80,20)(0,0)
    \Vertex(0,0){2}
    \Line(0,0)(50,0)
    \Vertex(50,0){2}
    \Text(20,-10)[]{$p$}
    \end{picture}
\hspace{1cm} S\left(p\right)=\frac{i}{\PS-m+i\varepsilon}.
\end{eqnarray}
{\it ii) Photon Propagator in the Feynman Gauge}
\begin{eqnarray}\label{FX1b}
\SetScale{1}
    \begin{picture}(80,20)(0,0)
    \Vertex(0,0){2}
    \Photon(0,0)(50,0){3}{6}
    \Vertex(50,0){2}
    \Text(20,-10)[]{$k$}
    \Text(0,-10)[]{$\mu$}
    \Text(40,-10)[]{$\nu$}
    \end{picture}
\hspace{1cm} D_{\mu\nu}\left(k\right)=\frac{-ig_{\mu\nu}}{k^{2}}.
\end{eqnarray}
{\it iii) Ghost Propagator}
\begin{eqnarray}\label{FX1c}
\SetScale{1}
    \begin{picture}(80,20)(0,0)
    \Vertex(0,0){2}
    \DashLine(0,0)(50,0){2}
    \Vertex(50,0){2}
    \Text(20,-10)[]{$p$}
    \end{picture}
\hspace{1cm} G\left(p\right)=\frac{i}{p^{2}+i\varepsilon}.
\end{eqnarray}
{\it iv) The Vertex of Two Fermions and One Gauge Field}
\vskip0.2cm
\begin{eqnarray}\label{FX1d}
\SetScale{1}
    \begin{picture}(50,20)(0,0)
    \Vertex(0,0){2}
    \ArrowArc(0,0)(20,250,300)
    \Photon(0,0)(0,20){2}{4}
    \LongArrow(5,18)(5,12)
    \ArrowLine(-20,-20)(0,0)
    \ArrowLine(20,-20)(0,0)
    \Text(20,15)[]{$k_{1},\mu$}
    \Text(-25,-15)[]{$p_{1}$}
    \Text(25,-15)[]{$p_{2}$}
    \end{picture}
\hspace{0.5cm}
V_{\mu}\left(p_{1},p_{2};k_{1}\right)=\left(2\pi\right)^{4}\delta^{4}\left(p_{1}+p_{2}+k_{1}\right)2g\gamma_{\mu}\
\sin\left(p_{1}\times p_{2}\right).
\end{eqnarray}
\vskip0.5cm
\par\noindent
{\it v) The Vertex of Two Ghost Fields and One Gauge Field}
\vskip0.2cm
\begin{eqnarray}\label{FX1e}
\SetScale{1}
    \begin{picture}(50,20)(0,0)
    \Vertex(0,0){2}
    \ArrowArc(0,0)(20,250,300)
    \Photon(0,0)(0,20){2}{4}
    \LongArrow(5,18)(5,12)
    \DashLine(-20,-20)(0,0){2}
    \DashLine(20,-20)(0,0){2}
    \Text(20,15)[]{$k_{1},\mu$}
    \Text(-25,-15)[]{$p_{1}$}
    \Text(25,-15)[]{$p_{2}$}
    \end{picture}
\hspace{0.5cm}
G_{\mu}\left(p_{1},p_{2};k_{1}\right)=\left(2\pi\right)^{4}\delta^{4}\left(p_{1}+p_{2}+k_{1}\right)2g\
(p_{2})_{\mu}\ \sin\left(p_{1}\times p_{2}\right).
\end{eqnarray}
\vskip0.5cm
\par\noindent
{\it vi) Three-Photon Vertex} \vskip0.2cm
\begin{eqnarray}\label{FX1f}
\SetScale{1}
    \begin{picture}(50,20)(0,0)
    \Vertex(0,0){2}
    \Photon(0,0)(0,20){2}{4}
    \LongArrow(-10,20)(-10,12)
    \LongArrow(-25,-15)(-20,-8)
    \LongArrow(25,-15)(20,-8)
    \Photon(-20,-20)(0,0){2}{4}
    \Photon(20,-20)(0,0){2}{4}
    \Text(20,15)[]{$p_{1},\mu_{1}$}
    \Text(-40,-20)[]{$p_{2},\mu_{2}$}
    \Text(40,-20)[]{$p_{3},\mu_{3}$}
    \end{picture}
\hspace{0.2cm}
\lefteqn{W_{\mu_{1}\mu_{2}\mu_{3}}\left(p_{1},p_{2},p_{3}\right)=-\left(2\pi\right)^{4}\delta^{4}\left(p_{1}+p_{2}+p_{3}\right)2g   \sin\left(p_{1}\times p_{2}\right) } \nonumber\\
&&\hspace{0.cm}\times \bigg[
g_{\mu_{1}\mu_{2}}\left(p_{1}-p_{2}\right)_{\mu_{3}} +
g_{\mu_{1}\mu_{3}}\left(p_{3}-p_{1}\right)_{\mu_{2}}+
g_{\mu_{3}\mu_{2}}\left(p_{2}-p_{3}\right)_{\mu_{1}} \bigg].
\end{eqnarray}
{\it vii) Four-Photon Vertex} \vskip0.2cm
\begin{eqnarray}\label{FX1g}
\SetScale{1}
    \begin{picture}(50,20)(0,0)
    \Vertex(0,0){2}
    \LongArrow(25,15)(20,8)
    \LongArrow(-25,15)(-20,8)
    \LongArrow(-25,-15)(-20,-8)
    \LongArrow(25,-15)(20,-8)
    \Photon(-20,20)(0,0){2}{4}
    \Photon(20,20)(0,0){2}{4}
    \Photon(-20,-20)(0,0){2}{4}
    \Photon(20,-20)(0,0){2}{4}
    \Text(-40,20)[]{$p_{1},\mu_{1}$}
    \Text(40,20)[]{$p_{4},\mu_{4}$}
    \Text(-40,-20)[]{$p_{2},\mu_{2}$}
    \Text(40,-20)[]{$p_{3},\mu_{3}$}
    \end{picture}
\hspace{0.2cm}
\lefteqn{W_{\mu_{1}\mu_{2}\mu_{3}\mu_{4}}\left(p_{1},p_{2},p_{3},p_{4}\right)=-4ig^2\left(2\pi\right)^{4}\delta^{4}\left(p_{1}+p_{2}+p_{3}\right)}\nonumber\\
&&\times\bigg[(g^{\mu_{1}\mu_{3}}g^{\mu_{2}\mu_{4}}-g^{\mu_{1}\mu_{4}}g^{\mu_{2}\mu_{3}})\sin(p_{1}\times p_{2})\sin(p_{3}\times p_{4})\nonumber\\
&&\ \ +(g^{\mu_{1}\mu_{4}}g^{\mu_{2}\mu_{3}}-g^{\mu_{1}\mu_{2}}g^{\mu_{3}\mu_{4}})\sin(p_{3}\times p_{1})\sin(p_{2}\times p_{4})\nonumber\\
&&\ \
+(g^{\mu_{1}\mu_{2}}g^{\mu_{3}\mu_{4}}-g^{\mu_{1}\mu_{3}}g^{\mu_{2}\mu_{4}})\sin(p_{1}\times
p_{4})\sin(p_{2}\times p_{3})\bigg].
\end{eqnarray}
Here we have used the notation: $p_{1}\times
p_{2}\equiv{\frac{1}{2}\theta_{\mu\nu}p_{1}^{\mu}p_{2}^{\nu}}$.
\section{Perturbative Calculation of the One-Loop $\beta$-Function}
\setcounter{equation}{0} In this section, we present the result of
the perturbative calculation of the one-loop $\beta$-function of
noncommutative U(1) gauge theory with adjoint matter field
coupling. But, let us briefly review the result of the one-loop
$\beta$-function of noncommutative U(1) with matter fields in the
fundamental representation, which is first presented in Ref.
\cite{hayakawa}.
\vskip0.5cm\par\noindent {\it i) One Loop $\beta$-Function of
NC-U(1) with Matter Fields in the Fundamental Representation}
\vskip0.5cm\par\noindent
The one-loop $\beta$-function receives contributions from one-loop
fermion self-energy, vacuum polarization tensor and vertex
function. In Figs. [1]-[3], all one-loop diagrams contributing to
these functions are presented. To calculate the one-loop
$\beta$-function, we would like to consider only the contributions
from the planar Feynman integrals corresponding to the fermion and
photon self-energy and vertex function. The nonplanar integrals
are assumed to be finite for finite noncommutativity parameter
$\theta$ and do not contribute to the $\beta$-function of the
theory \cite{hayakawa}. The $\beta$-function of the theory is
therefore not altered by the UV/IR mixing phenomena
\cite{minwalla}.
\par
As in the commutative QED, the $\beta$-function of the theory is
given by
\begin{eqnarray}\label{M7}
\beta(g)=\mu\frac{\partial}{\partial\mu}g(\mu).
\end{eqnarray}
In terms of the bare coupling constant $g_{0}$ and the
renormalization constants $Z_{i}, i=1,2,3$, the renormalized
coupling constant $g(\mu)$ is defined by:
\begin{eqnarray}\label{M8}
g_{0}=\mu^{-\epsilon/2}g(\mu)Z_{1}Z_{2}^{-1}Z_{3}^{-1/2},
\end{eqnarray}
with $\epsilon=4-D$.  In one-loop order, the renormalization
constant $Z_{1}$ receives contribution from one-loop vertex
function, whereas one-loop fermion and photon self-energy diagrams
contribute to $Z_{2}$ and $Z_{3}$, respectively. After regulating
the planar part of the corresponding one-loop Feynman integrals to
these diagrams, using {\it e.g.} dimensional regularization, the
one-loop contribution to $Z_{i}, i=1,2,3$ read
\begin{eqnarray}\label{M9}
Z_{1}^{{\small{Fund.}}}&=&1-\frac{g^2}{16\pi^2\epsilon},\nonumber\\
Z_{2}^{{\small{Fund.}}}&=&1-\frac{g^2}{8\pi^2\epsilon},\nonumber\\
Z_{3}^{{\small{Fund.}}}&=&1+\frac{g^2}{16\pi^2\epsilon}(\frac{20}{3}-\frac{8N_{f}}{3}).
\end{eqnarray}
Now using Eqs. (\ref{M8}) and (\ref{M9}), the one-loop
$\beta$-function of noncommutative QED with the matter fields in the
fundamental representation can be given by:
\begin{eqnarray}\label{M10}
\beta(g)\bigg|_{\small{1-loop}}^{{\small{Fund.}}}=-\frac{g^3}{16\pi^2}
\left(\frac{22}{3}-\frac{4N_{f}}{3}\right),
\end{eqnarray}
where $N_{f}$ is the number of the flavor degrees of freedom. For
$N_{f}<6$ the noncommutative QED with fundamental matters is
asymptotically free.
\vskip0.5cm\par\noindent {\it ii) One Loop $\beta$-Function of
NC-U(1) with Matter Fields in the Adjoint Representation}
\vskip0.5cm\par\noindent
Let us now consider the noncommutative QED with adjoint matter
fields coupling. Using the propagators and vertices of
noncommutative QED from Eqs. (\ref{FX1a})-(\ref{FX1g}), the
Feynman integral corresponding to the one-loop fermion self energy from
Fig. 1 can be given by
\begin{eqnarray}\label{M11}
-i\Sigma(p)=-4g^{2}\int
\frac{d^{D}k}{(2\pi)^{D}}\frac{\gamma^{\mu}(\KS+m)\gamma_{\mu}}{(k^{2}-m^{2})(p-k)^{2}}
\sin^{2}(\frac{p\times k}{2}).
\end{eqnarray}
Using now the relation $\sin^{2}x=\frac{1}{2}(1-\cos 2x)$, the
above integral can be separated into a planar and a nonplanar part.
The planar part reads:
\begin{eqnarray}\label{M12}
-i\Sigma_{planar}=-2g^{2}\int
\frac{d^{D}k}{(2\pi)^{D}}\frac{\gamma^{\mu}(\KS+m)\gamma_{\mu}}{(k^{2}-m^{2})(p-k)^{2}}.
\end{eqnarray}
It is exactly twice the value of the one-loop fermion self energy
from commutative QED. The wave function renormalization constant
$Z_{2}$ is then given by:
\begin{eqnarray}\label{M13}
Z_{2}^{Adj.}=1-\frac{g^{2}}{4\pi^{2}}\frac{1}{(4-D)}.
\end{eqnarray}
The one-loop photon self energy receives contributions from four
diagrams (2.a)-(2.d) from Fig. 2. It turns out that diagrams
(2.a)-(2.c) have both planar and nonplanar parts, whereas diagram
(2.d) has only a finite nonplanar part. We will consider only the
planar parts of the corresponding Feynman integrals, because they
are the only relevant parts to calculate the one-loop
$\beta$-function of the theory. The planar parts of the diagrams
(2.a)-(2.c) are given by:
\begin{eqnarray}\label{M14}
i\Pi^{\mu\nu}_{a-planar}&=& -2g^{2}\int
\frac{d^{D}k}{(2\pi)^{D}}\frac{\mbox{Tr}(\gamma^{\mu}
(\KS-\PS+m)\gamma^{\nu}(\KS+m))}{(k^{2}-m^{2})(k-p)^{2}},\nonumber\\
i\Pi^{\mu\nu}_{b-planar}&=&-2g^{2}\int\frac{d^{D}k}{(2\pi)^{D}}\frac{k^{\mu}k^{\nu}-k^{\mu}p^{\nu}}{k^{2}(k-p)^{2}},\nonumber\\
i\Pi^{\mu\nu}_{c-planar}&=&+2g^{2}\int\frac{d^{D}k}{(2\pi)^{D}}\frac{(D-6)p^{\mu}p^{\nu}-(2D-3)(k^{\mu}p^{\nu}+k^{\nu}p^{\mu}-2k^{\mu}k^{\nu})+g^{\mu\nu}(5p^{2}-2p.k+2k^{2})}{k^{2}(k-p)^{2}}.\nonumber\\
\end{eqnarray}
Using dimensional regularization, it is possible to isolate the
UV-divergent part of the above integrals, which is given by
\begin{eqnarray}\label{M15}
i\bigg[\Pi^{\mu\nu}_{a-planar}+\Pi^{\mu\nu}_{b-planar}+\Pi^{\mu\nu}_{c-planar}\bigg]_{\mbox{div.}}=
\frac{ig^{2}}{16\pi^{2}}(p^{\mu}p^{\nu}-g^{\mu\nu}p^{2})
(\frac{-20}{3}+\frac{16N_{f}}{3})\frac{1}{(4-D)}.
\end{eqnarray}
The wave function renormalization constant $Z_{3}$ can then be
obtained by making use of standard field theoretical definitions
and reads:
\begin{eqnarray}\label{M16}
Z_{3}^{Adj.}=1+\frac{g^{2}}{16\pi^{2}}(\frac{20}{3}-\frac{16N_{f}}{3})\frac{1}{(4-D)}.
\end{eqnarray}
As next, let us consider the one-loop correction to the vertex
function, which receives contribution from two diagrams (3.a) and
(3.b) from Fig. 3. The corresponding Feynman integrals to diagram
(3.a) is given by:
\begin{eqnarray}\label{M17}
\lefteqn{\hspace{-1cm}2g\sin(\frac{p\times
p^{\prime}}{2})\Gamma_{a}^{\mu}(p, p', q=p'-p, m)=}\nonumber\\
&&\hspace{-1cm}+8ig^{3}\int
\frac{d^{D}k}{(2\pi)^{D}}{\frac{\gamma^{\alpha}(\KS+m)\gamma^{\mu}(\KS+\QS+m)\gamma_{\alpha}}{(k^{2}-m^{2})((k+q)^{2}-m^{2})(p-k)^{2}}}\sin(\frac{p\times
k}{2})\sin(\frac{(q+k)\times p^{\prime}}{2})\sin(\frac{k\times
q}{2}),
\end{eqnarray}
and the one-loop Feynman integral corresponding to diagram (3.b)
reads
\begin{eqnarray}\label{M18}
\lefteqn{\hspace{-0.5cm}2g\sin(\frac{p\times
p^{\prime}}{2})\Gamma_{b}^{\mu}(p,p',q=p'-p,m)=}\nonumber\\
&&-8ig^{3}\int
\frac{d^{D}k}{(2\pi)^{D}}\frac{\gamma_{\alpha}(\KS+m)\gamma_{\eta}[(p-2p^{\prime}+k)^{\alpha}g^{\mu\eta}+(p+p^{\prime}-2k)^{\mu}g^{\beta\eta}+(k+p^{\prime}-2p)^{\eta}g^{\mu\beta}]}{(k^{2}-m^{2})(p-k)^{2}(p^{\prime}-k)^{2}}\nonumber\\
&&\hspace{4cm}\times\sin(\frac{p\times
k}{2})\sin(\frac{k\times
p^{\prime}}{2})\sin(\frac{(k-p^{\prime})\times q}{2}).
\end{eqnarray}
Using again some standard trigonometrical relations, the planar
and the nonplanar parts of these integrals can be separated. The
planar parts are given by
\begin{eqnarray}\label{M19}
\Gamma^{\mu}_{a-planar}(p,p',q=p'-p,m)=g^{2}\int
\frac{d^{D}k}{(2\pi)^{D}}{\frac{\gamma^{\alpha}(\KS+m)\gamma^{\mu}(\KS+\QS+m)\gamma_{\alpha}}{(k^{2}-m^{2})((k+q)^{2}-m^{2})(p-k)^{2}}},
\end{eqnarray}
for diagram (3.a) and
\begin{eqnarray}\label{M20}
\lefteqn{\Gamma_{b-planar}^{\mu}(p,p',q=p'-p,m)=}\nonumber\\
&&ig^{2}\int
\frac{d^{D}k}{(2\pi)^{D}}\frac{\gamma_{\alpha}(\KS+m)\gamma_{\eta}
[(p-2p^{\prime}+k)^{\alpha}g^{\mu\eta}+
(p+p^{\prime}-2k)^{\mu}
g^{\beta\eta}+(k+p^{\prime}-2p)^{\eta}g^{\mu\beta}]}{(k^{2}-m^{2})(p-k)^{2}
(p^{\prime}-k)^{2}},\nonumber\\
\end{eqnarray}
for diagram (3.b).  Standard dimensional regularization is used,
as next, to obtain the divergent part of these two Feynman
integrals, which is given by
\begin{eqnarray}\label{M21}
\bigg[\Gamma^{\mu}_{a-planar}+\Gamma_{b-planar}^{\mu}\bigg]_{\mbox{div.}}=
-\frac{8g^{2}\gamma^{\mu}}{16\pi^{2}}\frac{1}{(4-D)}.
\end{eqnarray}
The scaling factor $Z_{1}$ is therefore given by
\begin{eqnarray}\label{M22}
Z_{1}^{{\small{Adj.}}}=1-\frac{8g^2}{16\pi^2}\frac{1}{(4-D)}.
\end{eqnarray}
Using now  Eqs. (\ref{M13}), (\ref{M16}) and (\ref{M22}) and the
relations (\ref{M8}) as well as (\ref{M7}), we obtain the one-loop
$\beta$-function of noncommutative QED with matter fields in the
adjoint representation
\begin{eqnarray}\label{M23}
\beta(g)\bigg|_{\small{1-loop}}^{{\small{Adj.}}}
=-\frac{g^3}{16\pi^2}\left(\frac{22}{3}-\frac{8N_{f}}{3}\right).
\end{eqnarray}
As it turns out, the noncommutative QED with adjoint matters is asymptotically
free for $N_{f}<3$. In the next section, the conformal anomaly of
the noncommutative U(1) gauge theory with adjoint matters will be calculated
explicitly.
\section{The Conformal Anomaly and the One-Loop $\beta$-Function}
\setcounter{equation}{0} 
In this section, the conformal anomaly of
noncommutative QED with matter fields in the adjoint
representation will be calculated using the Fujikawa's path
integral method \cite{fujikawa}. We will first introduce the
background field method to obtain the relevant path integral for
the effective action. Using this path integral, we will then
calculate the contributions of the matter, ghost and gauge fields
to the conformal anomaly, separately. The conformal anomaly of
noncommutative QED with adjoint matter field coupling will be
calculated eventually. Using the conformal anomaly, it is then
possible to obtain the one-loop $\beta$-function of the theory. Its value 
coincides with the result which we have obtained in the previous
section in the framework of perturbative expansion [see Eq. (\ref{M23})].
\subsection{The Background Field Method and the One-Loop Effective Action}
Let us first redefine the gauge field $A_{\mu}$ by $A_{\mu}\to
\frac{1}{g}A_{\mu}$. In the background field method, the
gauge field must be separated into two parts:
\begin{eqnarray}\label{M24}
A_{\mu}(x)=B_{\mu}(x)+a_{\mu}(x),
\end{eqnarray}
with $B_{\mu}$ as the classical part and  $a_{\mu}$ the quantum
fluctuating part. In terms of  $a_{\mu}$, the field strength
tensor from Eq. (\ref{M6}) is then given by:
\begin{eqnarray}\label{M25}
F_{\mu\nu}=F_{\mu\nu}[B]+D_{\mu}[B]a_{\nu}-D_{\nu}[B]a_{\mu}-i[a_{\mu},a_{\nu}]_{\star},
\end{eqnarray}
where $F_{\mu\nu}[B]$ and $D_{\mu}[B]$ are defined by
\begin{eqnarray}\label{M26}
F_{\mu\nu}[B]&\equiv&\partial_{\mu}B_{\nu}-\partial_{\nu}B_{\mu}
-i[B_{\mu},B_{\nu}]_{\star},\nonumber\\
D_{\mu}[B]a_{\nu}&\equiv&\partial_{\mu}a_{\nu}-i[B_{\mu},a_{\nu}]_{\star}.
\end{eqnarray}
Using these definitions, the pure gauge part of the action is
given by
\begin{eqnarray}\label{M27}
S_{gauge}[a_{\mu},B_{\mu}]&=&-\frac{1}{4g^{2}}\int d^{4}x
\left\{F_{\mu\nu}^{}[B]\star
F^{\mu\nu}[B]+4ia^{\mu}\star[F_{\mu\nu}[B],a^{\nu}]_{\star}\right.\nonumber\\
&&-\left.2a_{\mu}\star
D_{\nu}[B]D^{\nu}[B]a^{\mu}-2D_{\nu}[B]a^{\nu}\star
D_{\mu}[B]a^{\mu}+O(a_{\mu}^{3},a_{\mu}^{4})\right\}.
\end{eqnarray}
Like in the commutative gauge theory, the linear terms in
$a_{\mu}$ vanish by making use of the classical equation of motion
for $B_{\mu}$. The cubic and quadratic terms in $a_{\mu}$ can also
be neglected for small quantum fluctuations $a_{\mu}$. For constant
$B_{\mu}$, this action is invariant under the following
infinitesimal transformation of $a_{\mu}(x)$
\begin{eqnarray}\label{M28}
\delta a_{\mu}(x)=D_{\mu}[B]\alpha(x)-i[a_{\mu}(x),\alpha(x)]_{\star},
\end{eqnarray}
where $\alpha(x)$ is an arbitrary function of $x$. Here, like in
the commutative non-Abelian gauge theory, the gauge must be fixed.
The gauge fixing and the Fadeev-Popov parts of the action can then
be given by:
\begin{eqnarray}\label{M29}
S_{GF+FP}[a_{\mu},c,\bar{c}]=-\frac{1}{2g^{2}}\int
d^{4}x\left\{D_{\mu}[B]^{}a^{\mu}\star
D_{\nu}[B]a^{\nu}-i\bar{c}\star
D^{\mu}[B]\left(D_{\mu}[B]c-i[a_{\mu},c]_{\star}\right)\right\}.
\end{eqnarray}
In terms of the background field $B_{\mu}$, the matter field part
of the action reads
\begin{eqnarray}\label{M30}
S_{matter}[\psi,\bar{\psi}]=\int
d^{4}x\ \bar{\psi}(x)\star(i\DS[B]-m)\psi(x),
\end{eqnarray}
with
\begin{eqnarray}\label{M32}
D_{\mu}[B]\psi(x)=\partial_{\mu}\psi(x)-i[B_{\mu}(x),\psi(x)]_{\star}.
\end{eqnarray}
The resulting total action, is invariant under the following
infinitesimal transformations:
\begin{eqnarray}\label{M33}
\delta B_{\mu}&=&\partial_{\mu}\alpha(x)
-i[B_{\mu}(x),\alpha(x)]_{\star},\hspace{.5cm} \delta
a_{\mu}=i[a_{\mu}(x),\alpha(x)]_{\star} \nonumber\\
\delta c&=&i[c(x),\alpha(x)]_{\star},\hspace{2.7cm}\delta
\bar{c}=i[\bar{c}(x),\alpha(x)]_{\star},\nonumber\\
\delta\psi&=&-ig\alpha(x)\star\psi(x),\hspace{2.2cm}\delta\bar{\psi}=
ig\bar{\psi}(x)\star\alpha(x).
\end{eqnarray}
The effective action of the noncommutative QED with  matter fields
in the adjoint representation is therefore given by:
\begin{eqnarray}\label{M35}
\exp(-W[B])&=& \int D\psi\  D\bar{\psi}\ Da_{\mu}\ Dc\ D\bar{c}\
\exp(S^{gauge}_{quad}[a_{\mu}]+S^{ghost}_{quad}[c,\bar{c}]+S_{matter}[\psi,\bar{\psi}]),
\end{eqnarray}
with
\begin{eqnarray}\label{M36}
S^{gauge}_{quad}[a_{\mu}]&=&\frac{1}{2g^2}\int
d^{4}x\ \left(a_{\mu}(x)\star
D_{\nu}[B]D^{\nu}[B]a^{\mu}(x)-2ia_{\mu}(x)\star[F^{\mu\nu}[B],a_{\nu}(x)]_{\star}\right),
\end{eqnarray}
and
\begin{eqnarray}\label{M37}
S^{ghost}_{quad}[c,\bar{c}]&=&\frac{i}{4g^{2}}\int d^{4}x\
\bar{c}(x)\star D_{\mu}[B]D^{\mu}[B]c(x).
\end{eqnarray}
Further, $S_{matter}[\psi,\bar{\psi}]$ is given in Eq.
(\ref{M30}).
\subsection{The Conformal Anomaly of NC-QED}
As it is mentioned in Sect. 1, even in the massless limit, the
scale invariance of the noncommutative QED with matter fields in
the adjoint representation is broken explicitly due to the
noncommutativity parameter $\theta$. The theory is at the same
time anomalous under the global conformal transformations of
matter fields $\psi$ and $\bar{\psi}$
\begin{eqnarray}\label{M38}
\psi(x)\rightarrow\psi^{\prime}(x)=\exp(-\frac{1}{2}\alpha)\psi(x),\hspace{1cm}\mbox{and}\hspace{1cm}
\bar{\psi}(x)\rightarrow\bar{\psi}^{\prime}(x)=\exp(-\frac{1}{2}\alpha)\bar{\psi}(x),
\end{eqnarray}
gauge fields $a_{\mu}$ as well as ghost fields $c(x)$
\begin{eqnarray}\label{M39}
a_{\mu}(x)\rightarrow a^{\prime}_{\mu}(x)&=&e^{-\alpha}a_{\mu}(x),\nonumber\\
c(x)\to c^{\prime}(x)&=&e^{-2\alpha}c(x),
\hspace{1cm}\mbox{and}\hspace{1cm}\bar{c}\to
\bar{c}^{\prime}(x)=\bar{c}(x).
\end{eqnarray}
In this section, we intend to calculate the conformal anomaly of
noncommutative QED with adjoint matters arising from these global
transformations. The contributions of matter fields, gauge and
ghost fields to the conformal anomaly will be calculated
separately. We then
calculate the corresponding one-loop $\beta$-function of the
theory and will compare our result with the result from one-loop
perturbative calculation in the previous section.
\subsubsection*{The Contribution of the Matter Fields}
Let us assume that the Dirac operator $\DS\hspace{.1cm}[B]$ has a
complete set of eigenfunctions $\varphi_{n}(x)$ satisfying the
eigenvalue equation:
\begin{eqnarray}\label{M40}
\DS\hspace{.1cm}[B]\varphi_{n}(x)=\lambda_{n}\varphi_{n}(x),
\end{eqnarray}
with $\lambda_{n}$ the eigenvalues corresponding to the
eigenfunctions $\{\varphi_{n}(x)\}$. The orthonormality and
completeness relations for the set $\{\varphi_{n}(x)\}$ are defined
formally by the noncommutative $\star$-product and read
\begin{eqnarray}\label{M41}
\int
d^{4}x\varphi^{\dag}_{m}(x)\star\varphi_{n}(x)=\delta_{nm},\hspace{1cm}\sum_{n}\varphi^{\dag}_{n}(x)
\star\varphi_{n}(y)=\delta^{4}(x-y).
\end{eqnarray}
The matter fields can therefore be written as a sum of these
eigenfunctions
\begin{eqnarray}\label{M42}
\psi(x)=\sum_{n} a_{n}\varphi_{n}(x)
,\hspace{1cm}\bar{\psi}(x)=\sum_{m} \varphi^{\dag}_{m}(x)
\bar{b}_{m},
\end{eqnarray}
with the coefficients $a_{n}$ and $\bar{b}_{m}$ as Grassmannian
numbers. Let us now consider the matter field part of the
effective action
\begin{eqnarray}\label{M43}
\int D\psi\  D\bar{\psi}\ Da_{\mu}\ Dc\ D\bar{c}\
e^{S_{matter}[\psi,\bar{\psi}]},
\end{eqnarray}
with $S_{matter}[\psi,\bar{\psi}]$ given in Eq. (\ref{M30}). Under
infinitesimal conformal transformations of matter fields $\psi\to
\psi'=(1-\frac{\alpha}{2})\psi$ and $\bar{\psi}\to
\bar{\psi}'=(1-\frac{\alpha}{2})\bar{\psi}$ the fermionic measure
of the above action transforms as:
\begin{eqnarray}\label{M44}
D\psi D\bar{\psi}\rightarrow
D\psi^{\prime}D\bar{\psi}^{\prime}=J_{\psi}[\alpha ]D\psi
D\bar{\psi},
\end{eqnarray}
with the Jacobian
\begin{eqnarray}\label{M45}
J_{\psi}[\alpha ]&=&\det\bigg[\delta_{nm}-\frac{1}{2}\alpha\int
d^{4}x\ \varphi^{\dag}_{m}(x)\star\varphi_{n}(x)\bigg]^{-2}
=\exp\left(\alpha\sum_{n}\int
d^{4}x\ \varphi^{\dag}_{n}(x)\star\varphi_{n}(x)\right).
\end{eqnarray}
Here, we have used the relations $\det C=\exp[\mbox{Tr}\ln C]$ and
$\ln(1+\alpha)=\alpha+{\mathcal{O}}(\alpha^{2})$ for $\alpha<<1$.
The expression on the r.h.s. of Eq. (\ref{M45}) must be
regulated using a Gaussian damping factor. We arrive at
\begin{eqnarray}\label{M46}
J_{\psi}[\alpha]=\lim_{\epsilon\longrightarrow0}\hspace{.2cm}\exp\left(\alpha\sum_{n}\int\ 
d^{4}x\exp(-\epsilon\lambda_{n}^{2})\varphi^{\dag}_{n}(x)\star\varphi_{n}(x)\right).
\end{eqnarray}
Now using the eigenvalue equation (\ref{M40}) we get
\begin{eqnarray}\label{M47}
\lim_{\epsilon\longrightarrow0}\hspace{.2cm}\exp\left(\alpha\sum_{n}\int\ 
d^{4}x\exp_{\star}(-\epsilon\DS\hspace{.1cm}^{2})\varphi^{\dag}_{n}(x)\star\varphi_{n}(x)
\right) \equiv \exp\left(\alpha\int d^{4}x\ A_{\psi}(x)\right),
\end{eqnarray}
where $A_{\psi}$ is defined to be the contribution of matter
fields to the global conformal anomaly. After Fourier transforming
the eigenfunctions $\varphi_{n}$'s and using the completeness
relation of their Fourier transformed, we obtain:
\begin{eqnarray}\label{M48}
\int d^{4}x\
A_{\psi}(x)=\lim_{\epsilon\longrightarrow0}\hspace{.2cm}\int
d^{4}x\int \frac{d^{4}k}{(2\pi)^{4}}e^{-ikx}\star
\mbox{Tr}[\exp_{\star}(-\epsilon\DS\hspace{.1cm}^{2})]\star
e^{ikx}.
\end{eqnarray}
As next let us evaluate the exponent $\DS\hspace{0.1cm}^{2}$. Using the
definition of the covariant derivative from Eq. (\ref{M32}) in the
adjoint representation, we have:
\begin{eqnarray}\label{M49}
[D_{\mu}[B],D_{\nu}[B]]\chi(x)=
-i\int\frac{d^{4}k}{(2\pi)^{4}}F_{\mu\nu}(x;k)\tilde{\chi}(k)e^{ikx}.
\end{eqnarray}
Here $\chi(x)$ is an arbitrary function of $x$ and
\begin{eqnarray}\label{M50}
F_{\mu\nu}(x;k)\equiv 
-2i\int\frac{d^{4}p}{(2\pi)^{4}}\tilde{F}_{\mu\nu}(p)e^{ipx}\sin(\frac{p\times
k}{2}).
\end{eqnarray}
Choosing now $\chi(x)=[D_{\alpha},D_{\beta}]\varphi(x)$ we obtain
\begin{eqnarray}\label{M51}
\lefteqn{\left[D_{\mu}[B],D_{\nu}[B]\right]\left[D_{\alpha}[B],D_{\beta}[B]\right]\varphi(x)=}\nonumber\\
&&=
-4\int\frac{d^{4}p}{(2\pi)^{4}}\frac{d^{4}q}{(2\pi)^{4}}\tilde{F}_{\mu\nu}(p)
\tilde{F}_{\alpha\beta}(q)e^{i(p+q)x}\sin^{2}\left(\frac{p\times q
}{2}\right)\varphi(x).
\end{eqnarray}
Using further the relation $2\sin^{2}x=(1-\cos 2x)$ to separate the
planar from the nonplanar part of the above expression, we arrive
at
\begin{eqnarray}\label{M52}
\left[D_{\mu}[B],D_{\nu}[B]\right]\left[D_{\alpha}[B],D_{\beta}[B]\right]
\bigg|_{planar}=-2F_{\mu\nu}(x)F_{\alpha\beta}(x),
\end{eqnarray}
which leads automatically to
\begin{eqnarray}\label{M53}
[D_{\mu},D_{\nu}]\bigg|_{planar}=-i\sqrt{2}F_{\mu\nu}.
\end{eqnarray}
Now let us go  back to the equation (\ref{M48}) to evaluate $\DS\hspace{0.1cm}^{2}=D_{\mu}D_{\nu}\gamma_{\mu}\gamma_{\nu}$. After a rescaling of
the momentum $k_{\mu}\to \frac{k_{\mu}}{\sqrt{\epsilon}}$ with
$\epsilon$ the cutoff appearing in the damping factor from Eq.
(\ref{M48}), and using the relations
$\gamma_{\mu}\gamma_{\nu}=-\delta_{\mu\nu}+\sigma_{\mu\nu}$ with
$\sigma_{\mu\nu}=\frac{1}{2}[\gamma_{\mu},\gamma_{\nu}]$ and
$e^{-ikx}\star\partial_{\mu}e^{ikx}=\partial_{\mu}+ik_{\mu}$, we
arrive at
\begin{eqnarray}\label{M54}
\lefteqn{\int d^{4}x\
A_{\psi}(x)\bigg|^{Adj.}_{planar}=}\nonumber\\
&&=\lim_{\epsilon\longrightarrow0}\hspace{.2cm}\frac{1}{\epsilon^{2}}\int
d^{4}x\int \frac{d^{4}k}{(2\pi)^{4}}e^{k_{\mu}k^{\mu}}
\mbox{Tr}\bigg[\exp_{\star}\left(-2i\sqrt{\epsilon}k^{\mu}D_{\mu}-\epsilon
D_{\mu}D^{\mu}-\frac{i\sqrt{2}}{2}\epsilon\sigma_{\mu\nu}F^{\mu\nu}(x)\right)\bigg].
\end{eqnarray}
As next, the trace appearing on the r.h.s. of the above equation
must be expanded in the orders of $\epsilon$. Due to the factor
$\frac{1}{\epsilon^{2}}$ appearing before the integral, the
expansion must be taken up to the fourth order in $\epsilon$. All
higher orders in $\epsilon$ vanish in the limit $\epsilon\to 0$.
Making use of the relations $
\mbox{Tr}(\gamma^{\mu}\gamma^{\nu})=4\delta^{\mu\nu}$,
$\mbox{Tr}(\sigma^{\mu\nu})=0$, and
$\mbox{Tr}(\sigma^{\mu\nu}\sigma^{\alpha\beta})=
4(\delta^{\mu\beta}\delta^{\alpha\nu}-\delta^{\mu\alpha}\delta^{\beta\nu})$
and the integrals
\begin{eqnarray}\label{M55}
\int\frac{d^{4}k}{(2\pi)^{4}}e^{k^{\rho}k_{\rho}}[1,k^{\mu}k^{\nu},
k^{\mu}k^{\nu}k^{\alpha}k^{\beta}]= \frac{1}{(4\pi)^{2}}[1,\frac
{-\delta^{\mu\nu}}{2},\frac{1}{4}(\delta^{\mu\alpha}\delta^{\beta\nu}+\delta^{\mu\beta}
\delta^{\alpha\nu}+\delta^{\mu\nu}\delta^{\beta\alpha})],
\end{eqnarray}
the contribution of the adjoint matter fields to the global
anomaly of noncommutative QED with respect to the scale
transformations of matter fields from Eq. (\ref{M38}) is given
by
\begin{eqnarray}\label{M56}
\int d^{4}x\
A_{\psi}(x)\bigg|^{Adj.}_{planar}=\frac{1}{12\pi^{2}}\int
d^{4}x\ F^{\mu\nu}(x)\star F_{\mu\nu}(x).
\end{eqnarray}
\subsubsection*{The Contribution of the Ghost and Gauge Fields}
As next, the contributions of the gauge and ghost fields to the
conformal anomaly are to be calculated. After Eq. (\ref{M39}) the
infinitesimal global scale transformation of gauge and ghost
fields are given by:
\begin{eqnarray}\label{XX}
a_{\mu}(x)\to a'_{\mu}(x)=(1-\alpha)a_{\mu}(x),
\hspace{1cm}\mbox{and}\hspace{1cm} c(x)\to c'(x)=(1-2\alpha)c(x),
\end{eqnarray}
respectively.
\par
Let us now consider the eigenvalue equations of the operators
$D_{\nu}[B]D^{\nu}[B]-2i[F^{\mu\nu}[B], \cdot]_{\star}$ from the
gauge part of the effective action $S_{quad}^{gauge}$ [see Eq.
(\ref{M36})] and $D_{\nu}[B]D^{\nu}[B]$ from the ghost part of the
effective action $S_{quad}^{ghost}$ [see Eq. (\ref{M37})]
\begin{eqnarray}\label{M57}
D_{\nu}[B]D^{\nu}[B]V_{n}^{\mu}(x)-2i[F^{\mu\nu}[B](x),V_{n,\nu}(x)]_{\star}&=&
\xi_{n}V_{n}^{\mu}(x),\nonumber\\
D_{\nu}[B]D^{\nu}[B]S_{n}(x)&=&\zeta_{n}S_{n}(x),
\end{eqnarray}
where $\xi_{n}$ and $\zeta_{n}$ are the eigenvalues corresponding
to the eigenfunctions  $\{V_{n}^{\mu}(x)\}$ as well as
$\{S_{n}(x)\}$. Both sets of eigenfunctions build a complete and
orthonormal basis of the Hilbert space and can therefore be used
to describe the gauge and the ghost fields
\begin{eqnarray}\label{M58}
a_{\mu}(x)=\sum_{n} a_{n}V_{n,\mu}(x),\hspace{0.5cm}
c(x)=\sum_{n}b_{n}S_{n}(x),
\end{eqnarray}
where $a_{n}$ are c-number, whereas $b_{n}$ are Grassmannian
numbers. As next, let us consider the gauge and ghost part of the
effective action, which are given by
\begin{eqnarray}\label{M59}
\int D\psi\ D\bar{\psi}\ Da_{\mu} Dc\  D\bar{c}\
e^{S_{quad}^{gauge}[a_{\mu}]},\hspace{1cm}\mbox{and}\hspace{1cm}
\int D\psi\ D\bar{\psi}\ Da_{\mu} Dc\  D\bar{c}\
e^{S_{quad}^{ghost}[c,\bar{c}]},
\end{eqnarray}
with $S_{quad}^{gauge}[a_{\mu}]$ and $S_{quad}^{ghost}[c,\bar{c}]$
from Eq. (\ref{M36}) and (\ref{M37}), respectively. Under the
infinitesimal scale transformation, the measures of the above
integrals transform as:
\begin{eqnarray}\label{M60}
Da^{\prime}_{\mu}=J_{a}[\alpha]Da_{\mu},\hspace{1cm}\mbox{and}\hspace{1cm}
Dc^{\prime}\ D\bar{c}^{\prime}=J_{c}[\alpha]Dc\ D\bar{c},
\end{eqnarray}
with
\begin{eqnarray}\label{M61}
J_{a}[\alpha]&=&\det[\delta_{nm}+\alpha\int d^{4}x\
V_{m,\mu}(x)\star
V_{n}^{\mu}(x)]^{-1},\nonumber\\
J_{c}[\alpha]&=&\det[\delta_{nm}-2\alpha\int d^{4}x\ S_{m}(x)\star
S_{n}(x)]^{-1}.
\end{eqnarray}
As it turns out both Jacobians $J_{a}$ and $J_{c}$ are divergent
and have to be regulated using an ordinary exponential damping
factor. The contributions of the gauge and ghost fields to the
conformal anomaly, $A_{a}$ and $A_{c}$, are then defined by
\begin{eqnarray}\label{M62}
\int d^{4}x\ A_{a}(x)&\equiv& \lim_{\epsilon\longrightarrow
0}\left(-\sum_{n}\int d^{4}x\ V_{n,\mu}(x)\star\ e^{-\epsilon\xi_{n}}\
V_{n}^{\mu}(x)\right),\nonumber\\
 \int d^{4}x\ A_{c}(x)&\equiv& \lim_{\epsilon\longrightarrow
0}\left(2\sum_{n}\int d^{4}x\ e^{-\epsilon\zeta_{n}}S_{n}(x)\star
S_{n}(x)\right).
\end{eqnarray}
As next, the eigenvalue equations (\ref{M57}) must be used to
replace the eigenvalues $\xi_{n}$ and $\zeta_{n}$ by their
corresponding operators. After a rescaling $k_{\mu}\to
\frac{k_{\mu}}{\sqrt{\epsilon}}$ with $\epsilon$ as the same
cutoff appearing in the damping factors and using the same
algebraic relations leading from Eq. (\ref{M47}) to Eq.
(\ref{M54}) in the fermionic case, we arrive at:
\begin{eqnarray}\label{M63}
\lefteqn{\int d^{4}x\ A_{a}(x)=\nonumber}\\
&=&\lim_{\epsilon\longrightarrow 0}\frac{-1}{\epsilon^{2}}\int
d^{4}x\int
\frac{d^{4}k}{(2\pi)^{4}}e^{k^{\rho}k_{\rho}}\mbox{Tr}\bigg[\exp\left(-2i\sqrt{\epsilon}k^{\alpha}D_{\alpha}[B]\delta^{\mu\nu}-\epsilon
D^{2}[B]+2i\epsilon F^{\mu\nu}(x;k)\right)\bigg],
\end{eqnarray}
for the contribution of the gauge fields and
\begin{eqnarray}\label{M64}
\int d^{4}x\ A_{c}(x)&=&\lim_{\epsilon\longrightarrow
0}\frac{2}{\epsilon^{2}}\int d^{4}x
\int\frac{d^{4}k}{(2\pi)^{4}}e^{k^{\rho}k_{\rho}}\exp\left(-2i
\sqrt{\epsilon}k^{\mu}D^{\mu}-\epsilon
D_{\mu}[B]D^{\mu}[B]\right),
\end{eqnarray}
for the contribution of the ghost fields. The function
$F_{\mu\nu}(x;k)$ on the r.h.s. of the Eq. (\ref{M63}) is defined
in the Eq. (\ref{M50}). Going through the same algebraic
manipulations leading from Eq. (\ref{M54}) to Eq. (\ref{M56}) in
the fermionic case, we arrive at the contributions of the gauge
and ghost fields to the conformal anomaly:
\begin{eqnarray}\label{M65}
\int d^{4}x\  A_{a}(x)\bigg|_{planar}&=&-\frac{5}{24\pi^{2}}\int
d^{4}x\ F_{\mu\nu}(x)\star F^{\mu\nu}(x),\nonumber\\
\int d^{4}x\ A_{c}(x)\bigg|_{planar}&=&-\frac{1}{48\pi^{2}}\int
d^{4}x\ F^{\mu\nu}(x)\star F_{\mu\nu}(x).
\end{eqnarray}
Now, adding these contributions to the contribution of the fermionic fields from Eq. (\ref{M56}), the total conformal anomaly of the
noncommutative QED with matter fields in the adjoint
representation is then given by
\begin{eqnarray}\label{M66}
\int d^{4}x\  A(x)\bigg|^{Adj.}_{planar}&=&N_{f}\int
 d^{4}x\ A_{\psi}(x)\big|_{planar}+\int d^{4}x\ A_{a}(x)\big|_{planar}+
 \int d^{4}x\ A_{c}(x)\big|_{planar}\nonumber\\
 &=&\frac{1}{(4\pi)^{2}}(\frac{4}{3}N_{f}-\frac{11}{3})\int\
d^{4}x\ F^{\mu\nu}(x)\star F_{\mu\nu}(x).
\end{eqnarray}
Here, $N_{f}$ is the number of flavor degrees of freedom. In Ref.
\cite{nakajima} the conformal anomaly of noncommutative QED with
matter fields in the fundamental representation was calculated
using the same path integral method. It was given by
\begin{eqnarray}\label{M67}
\int d^{4}x\ A(x)\bigg|^{Fund.}_{planar}
=\frac{1}{(4\pi)^{2}}(\frac{2}{3}N_{f}-\frac{11}{3})\int\
d^{4}x\ F^{\mu\nu}(x)\star F_{\mu\nu}(x).
\end{eqnarray}
Comparing to the result from Eq. (\ref{M66}), only the
contribution to the matter fields to the scale anomaly differs
from the corresponding contribution to the anomaly of the
noncommutative $U(1)$ gauge theory with adjoint matter fields.
\subsection{One-loop $\beta$-Function of Noncommutative QED with
Adjoint Matters} 
As in the ordinary commutative case, in
noncommutative QED there is a relation between the scale anomaly
and the $\beta$-function of the theory which is given by
\begin{eqnarray}\label{M68}
\int d^{4}x\ A(x)=\frac{\beta(g)}{2g^{3}}\int d^{4}x\ 
F^{\mu\nu}(x)\star F_{\mu\nu}(x).
\end{eqnarray}
The function $A(x)$ on the l.h.s. of this equation is the scale
anomaly including the contributions of matter, gauge and ghost
fields. This is an exact relation and is satisfied
nonperturbatively for all orders of perturbative expansion.
\par
To calculate the one-loop $\beta$-function of the theory using Eq.
(\ref{M68}), it is enough to consider the conformal anomaly
arising from the quadratic terms in $a_{\mu}$ and neglecting all
higher order terms. Here, $a_{\mu}$ describes the quantum
fluctuations of the gauge fields from the constant background
field $B_{\mu}$.
\par
Using now the result from Eq. (\ref{M66}) and the relation
(\ref{M68}), the one-loop $\beta$-function of noncommutative QED
with adjoint matters can be calculated and reads:
\begin{eqnarray}\label{M69}
\beta(g)\bigg|^{Adj.}_{1-loop}=-\frac{g^{3}}{(4\pi)^{2}}(\frac{22}{3}-\frac{8}{3}N_{f}).
\end{eqnarray}
This result coincides with the one-loop $\beta$-function from Eq.
(\ref{M23}), which was found by performing an explicit
perturbative calculation up to one-loop order. In Ref.
\cite{nakajima}, the one-loop $\beta$-function of noncommutative
QED with fundamental matter fields was calculated using the same
relation (\ref{M68}) between the anomaly and the $\beta$-function
of the theory. It was given by the same perturbative result from
Eq. (\ref{M10}). 
\section{Conclusion}
In this paper, the one-loop $\beta$-function of noncommutative QED with adjoint matter fields  is calculated using two different methods. First, a perturbative analysis up to one-loop order is performed and the value of the one-loop $\beta$-function is determined. To obtain the renormalization constants $Z_{i}, i=1,2,3$, only the planar parts of the one-loop fermion- and photon self energy, as well as vertex function are evaluated using dimensional regularization. The nonplanar part of the corresponding one-loop Feynman integrals are assumed to be finite for finite noncommutativity parameter $\theta$. The value of the one-loop $\beta$-function is therefore not altered by the UV/IR mixing phenomena of noncommutative Field Theory. Since the $\beta$-function of the noncommutative QED does not include any noncommutativity parameter $\theta$, it is not possible to compare its value with the $\beta$-function of the ordinary commutative QED. The one-loop $\beta$-function of noncommutative QED is more comparable to the $\beta$-function of a commutative non-Abelian gauge theory, like QCD. We have found that noncommutative QED with matter fields in the adjoint representation is asymptotically free for $N_{f}<3$. As is known from Ref. \cite{hayakawa}, noncommutative QED with fundamental matter fields is asymptotically free for $N_{f}<6$. 
\par
The one-loop $\beta$-function of the theory is then determined by calculating the one-loop contributions to the conformal anomaly of the noncommutative QED with adjoint matter field coupling. It can be shown that classically, even in the massless limit, the scale invariance  of noncommutative Field Theories is explicitly broken. 
This is due to the noncommutativity of the space-time coordinates. In this paper, only the quantum corrections to the conformal anomaly are calculated up to one-loop order using the Fujikawa's path integral method. Considering only the planar contributions to the conformal anomaly, we have shown that the value of the one-loop $\beta$-function coincides in both methods.    
\section{Acknowledgment}
The authors thanks F. Ardalan, and H. Arfaei for useful
discussions.

\newpage
\section{Figures}
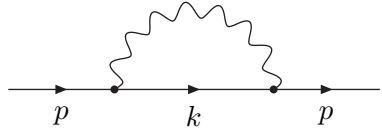
\begin{figure}[ht]
\begin{center}
\SetScale{1}
    \begin{picture}(50,20)(0,0)
    \Vertex(0,0){1.5}
    \Vertex(60,0){1.5}
    \ArrowLine(-40,0)(0,0)
    \ArrowLine(0,0)(60,0)
    \ArrowLine(60,0)(100,0)
    \Text(-20,-10)[]{$p$}
    \Text(30,-10)[]{$k$}
    \Text(80,-10)[]{$p$}
    \PhotonArc(30,0)(30,0,180){3}{8}
    \end{picture}
\end{center}
\vskip1.5cm \caption{One-loop contribution to the fermion self
energy.}
\end{figure}
\vskip1cm
\begin{figure}[ht]
\begin{center}
\SetScale{1}
    \begin{picture}(50,20)(0,0)
    \Vertex(70,0){1.5}
    \Vertex(110,0){1.5}
    \Photon(40,0)(70,0){3}{4}
    \Photon(110,0)(140,0){3}{4}
   \Text(55,-10)[]{$p$}
    \Text(90,30)[]{$k$}
    \Text(125,-10)[]{$p$}
    \PhotonArc(90,0)(20,0,360){3}{16}
     \Vertex(-50,0){1.5}
    \Vertex(-10,0){1.5}
    \Photon(-80,0)(-50,0){3}{4}
    \Photon(-10,0)(20,0){3}{4}
    \Text(-65,-10)[]{$p$}
    \Text(-30,30)[]{$k$}
    \Text(5,-10)[]{$p$}
    \DashArrowArc(-30,0)(20,0,360){2}
     \Vertex(-170,0){1.5}
    \Vertex(-130,0){1.5}
    \Photon(-200,0)(-170,0){3}{4}
    \Photon(-130,0)(-100,0){3}{4}
    \Text(-185,-10)[]{$p$}
    \Text(-150,30)[]{$k$}
    \Text(-115,-10)[]{$p$}
    \LongArrowArc(-150,0)(20,0,360)
      \Vertex(190,0){1.5}
    \Photon(160,0)(190,0){2.25}{4}
    \Photon(190,0)(220,0){2.25}{4}
    \Text(175,-10)[]{$p$}
   \Text(190,50)[]{$k$}
    \Text(205,-10)[]{$p$}
    \PhotonArc(190,20)(20,0,360){2.5}{12}
\Text(-90,0)[]{$+$} \Text(30,0)[]{$+$}\Text(150,0)[]{$+$}
\Text(190,-40)[]{$(d)$}\Text(-150,-40)[]{$(a)$}\Text(90,-40)[]{$(c)$}\Text(-30,-40)[]{$(b)$}
    \end{picture}
\end{center}
\vskip1.5cm \caption{One-loop contributions to the photon self
energy.}
\end{figure}
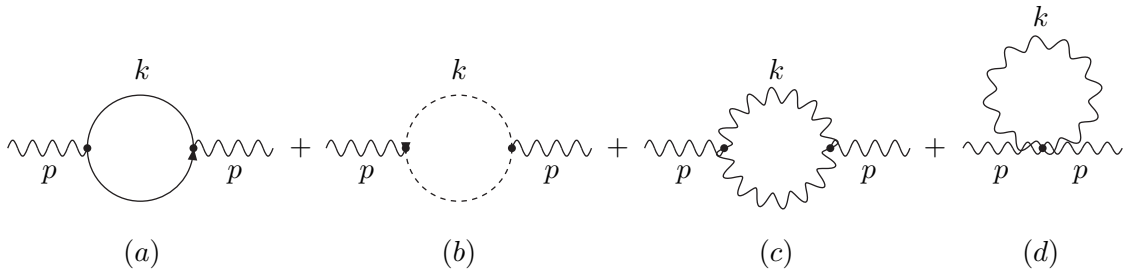
\vskip1cm
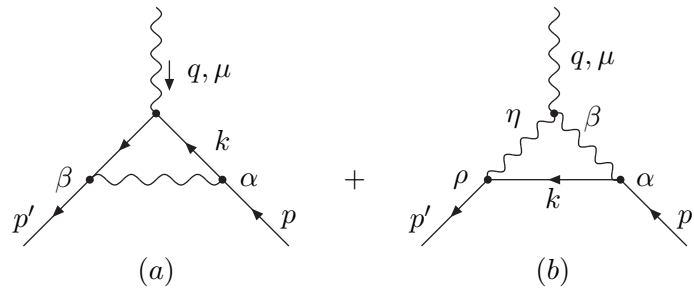
\begin{figure}[ht]
\begin{center}
\SetScale{1}
    \begin{picture}(40,40)(0,0)
    \Vertex(-50,0){1.5}
    \Photon(-50,0)(-50,40){2}{4}
    \Photon(-75,-25)(-25,-25){2}{4}
    \LongArrow(-45,20)(-45,10)
    \ArrowLine(-50,0)(-75,-25)
    \ArrowLine(-25,-25)(-50,0)
    \Vertex(-25,-25){1.5}
    \Vertex(-75,-25){1.5}
    \ArrowLine(-75,-25)(-100,-50)
    \ArrowLine(0,-50)(-25,-25)
    \Text(-30,15)[]{$q,\mu$}
    \Text(-100,-40)[]{$p'$}
    \Text(0,-40)[]{$p$}
    \Text(-25,-10)[]{$k$}
    \Text(-15,-25)[]{$\alpha$}
    \Text(-85,-25)[]{$\beta$}
    \Text(25,-25)[]{$+$}
    \Vertex(100,0){1.5}
    \Photon(100,0)(100,40){2}{4}
    \ArrowLine(125,-25)(75,-25)
    \Photon(100,0)(75,-25){2}{4}
    \Photon(125,-25)(100,0){2}{4}
    \Vertex(125,-25){1.5}
    \Vertex(75,-25){1.5}
    \ArrowLine(75,-25)(50,-50)
    \ArrowLine(150,-50)(125,-25)
    \Text(115,20)[]{$q,\mu$}
    \Text(50,-40)[]{$p'$}
    \Text(150,-40)[]{$p$}
    \Text(100,-32)[]{$k$}
    \Text(135,-25)[]{$\alpha$}
    \Text(65,-25)[]{$\rho$}
    \Text(115,-2)[]{$\beta$}
    \Text(85,-2)[]{$\eta$}
    \Text(-50,-60)[]{$(a)$}\Text(100,-60)[]{$(b)$}
    \end{picture}
\end{center}
\vskip2cm \caption{One-loop contributions to the vertex function.}
\end{figure}

\end{document}